\documentclass[12pt]{article}

\usepackage{setspace,graphicx,epstopdf,amsmath,amsfonts,amssymb,amsthm}
\usepackage{marginnote,datetime,enumitem,subfigure,rotating,fancyvrb}
\usepackage{hyperref,float}
\usepackage[longnamesfirst]{natbib}
\usepackage{mathtools}
\usepackage{caption}
\usepackage{float}
\usepackage{lscape}
\usdate

\usepackage[a4paper,bindingoffset=0.2in,%
left=0.8in,right=0.8in,top=1in,bottom=0.8in,%
            footskip=.35in]{geometry}
\usepackage[disable]{todonotes}

\makeatletter\let\chapter\@undefined\makeatother 

\usepackage{indentfirst} 
\usepackage{endnotes}    
\usepackage[labelfont=bf,labelsep=period]{caption}   
\usepackage{titlesec}
\begin{document}

\setlist{noitemsep}  
\onehalfspacing      
\renewcommand{\footnote}{\endnote}  

\author{MYKOLA PINCHUK\thanks{\rm Pinchuk Mykola is MSc student in the Sauder School of Business, UBC}}

\title{\Large \bf Zero-Leverage Puzzle}

\date{}              


\maketitle
\thispagestyle{empty}

\bigskip

\centerline{\bf ABSTRACT}

\begin{doublespace}  
  \noindent In this paper, I examine why some firms have zero leverage. I fail to find evidence that firms are unlevered because of managerial entrenchment since these firms do not have weaker corporate governance. I reject the hypothesis that firms become zero-leverage after prolonged periods of high market valuation, since before levering these firms do not suffer from declining valuations and continue to issue large amounts of equity. I find strong evidence in favor of the financial constraints explanation of the zero-leverage puzzle. Zero-leverage firms appear to be financially constrained using three different measures of financial constraints. I obtain mixed evidence on the financial flexibility hypothesis since all-equity firms increase investments and acquisitions after levering, but the probability of their levering decreased during the financial crisis. My results suggest that financial constraints are the first-order driver of zero-leverage behavior and are more important than less obvious explanations such as managerial entrenchment.  
\end{doublespace}

\medskip

\clearpage

\section{Introduction} \label{sec:Model}
Capital structure has always been at the heart of corporate finance literature. During the recent 60 years, it has been subject to extensive research, both theoretical and empirical. While this literature, without any doubt, contributed to our understanding of capital structure decisions, academic research produced as many answers as generated new questions.
A low-leverage puzzle is an example of such a challenge. It arose due to the inability of traditional capital structure theories to explain observable levels of leverage. This puzzle has been actively researched within the last 30 years and gave rise to both new theories of capital structure, such as dynamic trade-off theory (Fischer, Heinkel, Zechner, 1989), and new approaches to testing predictions of existing capital structure theories (Strebulaev, 2007). While extensive research offered numerous explanations of the low-leverage puzzle, with recent papers proposing models and estimation techniques, which generate predictions, consistent with data, a new phenomenon of zero-leverage firms was unraveled.
\paragraph{}
The zero-leverage puzzle is the most extreme form of a low-leverage puzzle. In order to identify whether a firm is underlevered, we need to determine optimal leverage using some capital structure model, which is subjective and can generate significantly different results. Some capital structure theories can produce high optimal leverage and consider a firm underlevered, while under another theory a firm will have optimal leverage. However, the zero-leverage firm is clearly underlevered, since no traditional capital structure model can predict zero leverage. Therefore, the main research interest in zero-leverage puzzle lies in the fact that zero-leverage puzzle is the most convenient way to study low-leverage puzzle. There are other advantages of using zero-leverage firms to examine low-leverage puzzle as well. Low leverage may often be caused by high equity values rather than result from the low-leverage policy. Using zero-leverage firms to study a low-leverage puzzle makes the identification of low-leverage firms independent of the leverage denominator, thus making results robust to equity value fluctuations. Another advantage of using a zero-leverage sample to study a low-leverage puzzle emerges in analyzing debt initiation decisions of zero-leverage firms, since any debt issuance by zero-leverage firms unambiguously means a change in financial policy, while for low-leverage firms it can be used to refinance previous debts. Studying determinants of zero-leverage policy can shed light on economic forces, which contribute to a firm`s decision to become unlevered, so it is important for better understanding capital structure decisions in general.
\paragraph{}
In this paper, I try to determine, which less traditional theories of capital structure can explain zero-leverage behavior. Previous research offered numerous explanations of this phenomenon, including managerial entrenchment, financial constraints, financial flexibility and high equity market valuations. In order to find a theoretical explanation, most consistent with the data, I test predictions, generated by each of the aforementioned explanations. 
Summary statistics show that zero-leverage firms compared to positive-leverage firms are smaller, younger, have higher cash reserves, and lower tangibility. Zero-leverage firms have higher valuations, grow faster, pay higher dividends, issue more equity, have more volatile cash flows and lower capital expenses. Results without matching indicate that zero-leverage firms are less profitable and have lower cash flow, while after matching zero-leverage firms appear more profitable and cash flow-rich. 
\paragraph{}
Analysis of the transitivity of zero-leverage behavior shows that firms are more likely to be unlevered at the time of going public (26\%) rather than at any other moment in time (19.5\%). The probabilities of firms transferring between zero-leverage and positive-leverage groups are 19.9\% and 3.46\% respectively, implying long-term distribution with only 14.8\% of zero-leverage firms. This means that to a significant extent, new firms, which usually lever up as time passes by, drive zero-leverage puzzle. 
\paragraph{}
Results of the analysis clearly reject explanations of zero-leverage behavior from the point of view of weak corporate governance, since firms with higher board independence and institutional ownership are more likely to be unlevered. Tests falsify a hypothesis, resulting from an explanation of zero-leverage puzzle with high equity market valuation, since prior to levering, zero-leverage firms do not suffer from declining market valuations and continue to issue large amounts of equity. 
\paragraph{}
I find strong evidence for both financial constraints and financial flexibility explanations. Zero-leverage firms are more financially constrained than positive-leverage firms using the SA-index, cash flow sensitivity of cash, and textual analysis of 10-K filings as measures of financial constraints. All-equity firms dramatically increase capital expenditures as well as acquisitions immediately after levering, consistent with the financial flexibility hypothesis. To discriminate between the financial constraints explanation and financial flexibility theory, I use the recent financial crisis as an exogenous shock to the economic environment. Since a crisis is unlikely to make previously financially constrained zero-leverage firms less financially constrained, the percentage of zero-leverage firms, deciding to issue debt, should decrease. The financial flexibility explanation predicts that zero-leverage firms will use their stockpiled debt capacity to issue debt in order to avoid cutting investments or acquiring distressed competitors. A dramatic decrease in the proportion of levering zero-leverage firms in 2009 provides clear and statistically significant evidence in favor of the financial constraints explanation and against the financial flexibility explanation.
\section{Hypotheses Development and Related Studies} \label{sec:Model}
\subsection{Conceptual Framework}
First mentions of zero-leverage firms logically emerged within research on the low-leverage puzzle (Graham, 2000). Minton and Wruck (2001) were the first to study the distinction between low-leverage firms and their high-levered counterparts. They find that low-leverage firms have high market-to-book ratios and utilize transitory low-leverage policies. To explain the financing decisions of low-leverage firms, their paper proposes a modified pecking order theory, implying that low-leverage firms stockpile their debt capacity to finance large discretionary expenditures in the future. While this paper does not address the issue of zero-leverage firms explicitly, it identifies basic stylized facts about low-leverage firms, which hold for zero-leverage firms as well, since the latter is a large subset of low-leverage firms. Furthermore, the paper proposed an original theoretical explanation of the behavior of low-leverage firms, which laid the foundation for financial flexibility theory. 
\paragraph{}
Literature on zero-leverage firms is young and limited in both scope and depth. Recent papers propose 5 different explanations of the zero-leverage phenomenon:
\begin{enumerate}
  \item Financial constraints.
  \item Financial flexibility.
  \item Underinvestment hypothesis.
  \item Managerial entrenchment
  \item High equity market valuation
\end{enumerate}
Explanation of zero-leverage decisions by financial constraints usually uses the theory, proposed by Faulkender and Petersen (2006). Their paper suggests that access to the debt market is a significant determinant of capital structure, since firms, unable to issue relatively cheap debt, will resort to equity financing. In the context of zero-leverage firms, these explanation is supported by several empirical research (Devos, 2012; Byoun, 2012; Bessler et al, 2012; ).
\paragraph{}
DeAngelo (2007) introduces the financial flexibility theory, though conceptually similar to the theory of Minton and Wruck (2001). It suggests that firms choose low leverage and large cash reserves in order to stockpile debt capacity and preserve their borrowing capacity that can be used to fund new large-scale investment opportunities in the future. Another variation of financial flexibility theory predicts that during an economic downturn underlevered firm will raise debt financing and use it for aggressive price competition. These explanations of the zero-leverage puzzle are supported by Minton and Wruck; Byoun, Besser et al.
\paragraph{}
The underinvestment hypothesis is built upon the theory of agency conflicts between stockholders and debtholders (Myers, 1977). This explanation suggests that for the management of highly levered firms, it can be optimal to forgo low-risk positive NPV projects since these projects are mainly financed by equityholders, while their proceeds are mainly paid to debtholders. For young fast-growing firms in highly competitive industries, underinvestment may be very costly, causing them to decrease leverage to avoid agency problems and underinvestment. Dang and Boroshko find evidence, supporting this view.
\paragraph{}
The managerial entrenchment hypothesis of the zero-leverage puzzle is based on previous research, suggesting an effect of CEO preferences on financing policy (Jensen 1986, Yermack 1997). Strebulaev (2013) extends this explanation by considering the attempts of managers with high stock ownership and underdiversified portfolios to decrease the risk of their portfolios by limiting their firms` leverage. While Strebulaev (2013) and Boroshko (2015) find evidence in favor of this explanation, Devos (2012) and Byoun (2012) report findings inconsistent with the managerial entrenchment explanation.
\paragraph{}
The high equity market valuation hypothesis suggests that the firms, whose stocks were significantly overpriced in the past, are more likely to be unlevered due to managerial attempts to take advantage of this overpricing. This explanation of the zero-leverage puzzle is supported by Byoun (2012). However, the findings of Lee and Moon are inconsistent with the equity market valuation hypothesis.
\subsection{Hypotheses Development}
\subsubsection{Depletion of post-IPO cash reserves}
The results above indicate that zero-leverage firms are usually small and young. This fact suggests a possible explanation of the zero-leverage puzzle, according to which small and young firms do not need any external financing, and thus do not issue debt or equity. Since the sample consists of public firms, firms enter the sample immediately after IPO, which lets them sufficiently increase cash holding. This explanation suggests that for some time after IPO firms have large cash reserves, sufficient to finance all their activities and investments, so they have no reason to issue any additional claims. As time passes by, their cash, raised after IPO, decreases, and firms start to resort to debt financing. This theory explains why zero-leverage firms are small and young and generate the following testable hypothesis.
\paragraph{}
\textbf{Hypothesis 1: As a ratio of current cash reserves to cash reserves immediately after IPO decreases, firms become more likely to issue debt.}

\subsubsection{Financial Constraints}
An important question as to the zero-leverage puzzle is whether firms choose not to issue debt or their zero leverage is not a result of their decision since they are not able to issue debt due to problems with access to debt financing. Financial constraints explanation suggests that unlevered firms would wish to issue debt to utilize tax and other advantages of debt, but do not have access to debt financing on acceptable terms. This explanation is consistent with the observation that zero-leverage firms are smaller and younger than positive-leverage firms since small and young firms are less likely to have established reputations on debt markets or relations with banks. Another possible problem with the access of small and young firms to debt financing is that these firms are considered riskier and are expected to pay higher interest to compensate for high risk. Financial constraints explanation generates testable predictions, formalized in hypothesis 2A.
\paragraph{}
\textbf{Hypothesis 2A: Zero-leverage firms are more likely than positive-leverage firms to be financially constrained.}
\paragraph{}
To test this hypothesis, I use different measures of financial constraints: financial constraints variables, derived from 10-K filings, SA--Index and cash flow sensitivity of cash.

\subsubsection{Financial Flexibility}
In contrast to financial constraints theory, financial flexibility theory suggests that zero leverage is result of management`s decision rather than condition, forced by circumstances. Zero-leverage firms are able to issue debt on acceptable terms, but they decide to temporarily abstain from it, in order to have more leeway in the future regarding financing options. This theory presents debt issuance decision as a trade-off between raising debt financing now and raising debt financing in the future. It the firm decide to issue debt now, it means that it will not be able to issue as much debt in the future as it would be able otherwise. Financial flexibility theory argues that zero-leverage policy is transitive, since firms are going back and forth between zero leverage and positive leverage. Firms use debt issuance to finance large-scale investment projects or acquisitions. Thus, under financial flexibility theory, firm waits until such projects arise and then undertakes debt issuance. Within next years, firm repays debt promptly (either because it is short-term debt or using early repayment) and becomes unlevered. Financial flexibility theory underpins hypothesis 3A. 
\paragraph{}
\textbf{Hypothesis 3A: After transition from zero leverage to positive leverage, firm increases investments or acquisitions.}
\paragraph{}
To test this hypothesis, I calculate capital expenditures, R\&D expenses and acquisitions before and after levering of zero-leverage firm.

\subsubsection{High market valuation}
Market valuation theory of capital structure considers current capital structure as a result of previous attempts of firm`s management to take advantage of market conditions. Under this theory, firm decides which claims to use by comparing conditions on equity market and debt market. Since debt is senior to equity, stock prices are more volatile than debt prices and it is equity market conditions, which matter most in firms` attempts to time the market. Thus market valuation theory predicts that zero-leverage firms had overvalued stocks in the past and their managements undertook large stock issuance in an attempt to take advantage of overvalued stocks. High market valuation explanation generates testable predictions, formalised in hypothesis 4A.
\paragraph{}
\textbf{Hypothesis 4A: Before the levering, zero-leverage firms have decreasing valuations and issue less equity than positive-leverage firms.}

\subsubsection{Managerial entrenchment}
Another explanation of zero-leverage phenomenon is based on the assumption that zero-leverage policy is suboptimal and is pursued by the managers due to managers` incentives. Explanation of exact reasons for managers to pursue zero-leverage policy may vary. For example, Strebulaev (2013) proposes modified theory by Yermack (1997), arguing that managers with high ownership in their firms are more likely to be underdiversified, since major part of their portfolio of financial assets have the same source of risk as their non-financial portfolio. Thus managers try to decrease their main source of risk – firm they are running. One of the methods of doing this is decreasing leverage, with zero-leverage policy being the most extreme form. 
\paragraph{}
Attempts to avoid disciplining effect of debt payment is another possible reason for suboptimal zero-leverage policy, imposed by managers. This notion is based on the assumption that managers prefer to have discretion over as much cash reserves as possible, partially in order to be able to waste cash on perquisites or suboptimal projects. I do not discriminate between these two possible explanation and use one single hypothesis to test this explanation, which does not depend on driving forces behind managerial entrenchment.
\paragraph{}
\textbf{Hypothesis 5: Firms with strong corporate governance are less likely to have zero leverage.}
\paragraph{}
Since zero-leverage policy is suboptimal, managers are destroying value by pursuing it. Therefore in firms with stronger corporate governance and external control it will be harder for mangers to pursue suboptimal zero-leverage policy.

\subsubsection{Multiple hypotheses: Financial constraints, Financial flexibility and High market valuation}
All hypotheses, proposed above, are not mutually exclusive, so, in case all of them are accepted, we need a way to discriminate between them. For this effect I use recent financial crisis as a setting, providing a source of exogenous shock for zero-leverage firms. Due to clearly exogeneous source of shock, I can establish causal effect of changes in economic environment on zero-leverage policy. 
\paragraph{}
Financial constraints explanation of zero-leverage puzzle entails that firms are zero-levered because they are unable to raise debt financing, which would increase their value. Financial and economic crisis in 2008 – 2009 was accompanied by deep crisis on the debt markets. Thus I argue that firms which were financially constrained in 2007, are highly unlikely to become financially unconstrained and obtain access to debt markets in 2008 -- 2009. On the other hand, many firms with access to debt market in 2007, lost it and became financially constrained in 2008 -- 2009. Financially constrained zero-leverage firms lack opportunities to raise funding from debt markets while, with regard to equity market drop in 2008, it is very unlikely that they decide to raise equity. Therefore, zero-leverage financially constrained firms have only internal source of cash, which suffers from plunge in consumption, and hence, in corporate sales. All mentioned facts mean that financially constrained firms are more fragile and find it harder to adapt to rapidly changing economic environment. Therefore, their performance is expected to be worse than those of unconstrained firms. 
\paragraph{}
\textbf{Hypothesis 2B: In 2008 - 2009 zero-leverage firms were less likely to issue debt and become positive-leverage firms. Performance of zero-leverage firms was worse than performance of positive-leverage firms.}
\paragraph{}
Financial flexibility explanation implies that firms choose zero-levrage policy in order to be able to issue debt in case of contingencies. Financial crisis is exactly type of event, during which firms, which choose zero-leverage policy to retain ability to issue debt, should take advangtage of this opportunity. These firms may use debt financing either to keep investments at previous level (it is well-known fact that firms cut investments during crisis) or to undertake cheap acquisitions of distressed competitors. So financial flexibility theory predicts that percentage of zero-levered firms, becoming positive-leverage firms, will increase. Zero-leverage firms will have higher investments than positive-leverage firms and undertake more acquisitions. These testable predictions are formalised into the following hypothesis.
\paragraph{}
\textbf{Hypothesis 3B: In 2008 and 2009 zero-leverage firms were more likely to issue debt and become positive-leverage firms. Investments and acquisitions of zero-leverage firms are higher than those of positive-leverage firms.}
\paragraph{}
High market valuation explanation of zero-leverage puzzle entails that zero-leverage firms became unlevered due to high prices of their shares and hence, massive equity issuance in the past, enough to cover all financing needs. Having regard to plunge in equity market in 2008, it is obvious that these firms can not count on equity issuance any more (even if their shares are still overpriced relative to competitors). Thus these firms will be forced to gradually start issuing debt. 
\paragraph{}
\textbf{Hypothesis 4B: In 2008 and 2009 zero-leverage firms were more likely to issue debt and become positive-leverage firms.}
\paragraph{}
Therefore, answer on the question whether percentage of levering zero-leverage firms increased or decreased in 2008 -- 2009 will let us discriminate between Hypothesis 2B and Hypotheses 3B with 4B.
\subsection{Related Empirical Studies}
Strebulaev and Yang (2013) conduct thorough empirical study of zero-leverage and ultra-low leverage firms (firms with leverage below 5\%). They find that this phenomenon is persistent in the long term and identify key differences between zero-leverage firms and their levered peers. Zero-leverage firms pay substantially higher dividends, have higher profitability and cash balances, pay more taxes and issue more equity. To emphasize inconsistency of zero-leverage financial policy with dominant theories of capital structure, they estimate increase in market value of zero-leverage firms by 7\% in case of levering up to a level of their matched proxies. Strebulaev and Yang explain existence of zero-leverage firms by managerial entrenchment. They find that firms with high CEO ownership and less independent boards are more likely to pursue zero-leverage policy. These findings and explanation by the managerial entrenchment are consistent with previous literature (Jensen, 1986; Yermack, 1997). \paragraph{}
Devos (2012) investigates zero-leverage firms and tests two alternative explanations using managerial entrenchment and financial constraints. He reports evidence that zero-leverage firms are young, small and have high cash balances. He does not find support for managerial entrenchment hypothesis and argues that zero-leverage firms are financially constrained, implying that their financial structure does not result from their choice. Since zero-levered firms lose market share in economic downturns, Devos argues that financial constraints explain zero-leverage policy better than financial flexibility theory.
\paragraph{}
Byoun (2012) fails to find evidence in favor of managerial entrenchment explanation and argues that zero-leverage firms are unlevered due to both financial constraints and high equity market valuations. He notes that high valuations of zero-leverage firms are likely to indicate overvaluation of their stock, which cause them to resort to equity financing. While these findings are in line with Devos (2012), in studying zero-leverage firms Byoun goes further and proposes to divide them into two groups on the basis of their characteristics and plausible explanations of their financial policy. He finds that most zero-leverage firms are small and young firms, which actively raise equity financing. Thus Devos argues that financing decisions of these firms are mainly governed by financial constraints. Another group of zero-leverage firms consists of large and profitable firms, which have positive cash flows and sufficient internal funds. While reporting detailed evidence on this group, he does not provide clear explanation for their financial decisions.
\paragraph{}
Numerous papers investigate zero-leverage phenomenon in international scope. Dang (2009) finds that zero-leverage phenomenon in the UK is even more frequent that in the US. He shows that characteristics of British zero-leverage firms are similar to those of zero-leverage firms in the US and finds evidence in favor of underinvestment hypothesis and dynamic trade-off theory to explain financial decisions of zero-leverage firms. His analysis rejects explanation of zero-leverage puzzle from pecking order theory and finds mixed evidence for financial flexibility theory.
\paragraph{}
Besser, Drobetz, Haller and Meier (2012) investigate international sample from Compustat Global and find the same characteristics of zero-leverage firms. They analyze recent upward trend in a fraction of zero-leverage firms and explain it by higher IPO activity together with observation that new public firms are more likely to be unlevered. A shift in industrial composition of stock market towards industries with higher fraction of all-equity firms is another plausible explanation. In explaining zero-leverage puzzle, they base their analysis on Byoun (2012) and extend his attempt to divide zero-leverage firms into two groups. They argue that majority of zero-leverage firms are unlevered due to financial constraints. These firms are usually small, young, do not generate enough cash flows and have small debt capacity. They remain unlevered for a long time and actively raise equity financing. Another group consists of large and profitable unlevered firms, which have sufficient debt capacity, but choose zero-leverage policy. Besser, Drobetz, Haller and Meier argue that financial flexibility is the main reason for their financial decision to abstain from debt and prove it by lower persistence of zero leverage policy of large unlevered firms.
\paragraph{}
Boroshko and Kokoreva (2015) study zero-leverage firms in Russia and Brazil and find support for managerial entrenchment explanation and underinvestment hypothesis. They report mixed evidence in favor of financial constraints hypothesis and argue that managerial entrenchment is the best explanation of zero-leverage puzzle in Russia and Brazil. Li and Gao (2015) investigate effect of zero-leverage policy on inefficient investment in China. They find that zero-leverage firms undertake more inefficient investment than levered firms and explain this effect by the lack of bank monitoring. Their evidence suggests that effect of zero-leverage policy on propensity to undertake inefficient investment is sufficiently lower for unlevered firms with strong external monitoring.
\paragraph{}
Lee and Moon (2011) examine stock returns of firms with persistent zero-leverage policy in the US. They find that these firms have significantly positive abnormal returns. This finding is remarkable, since it contradicts equity market valuation hypothesis explaining zero-leverage puzzle by firms` attempts to benefit from overvalued stocks. Therefore, equity market valuation hypothesis predicts that zero-leverage firms will have negative abnormal returns.
\paragraph{}
Lee and Moon (2015) extend their previous research by distinguishing between financially constrained zero-leverage firms and zero-leverage firms with high debt capacity. They find that both groups of unlevered firms have positive abnormal returns, with unconstrained firms generating significantly higher returns that their constrained counterparts. They suggest that this effect shows that financial decision to abstain from debt financing indicates disciplined and prudent management, which explains better performance of all-equity firms.

\section{Data and summary statistics} \label{sec:Model}

In this section, I describe the data and present summary statistics.
The sample is constructed from merged annual Compustat and CRSP databases for the period 1996--2015. Other sources of data include BoardEx, Compustat Execucomp, Thomson Reuters and 10-K SEC filings. In order to test managerial entrenchment hypothesis, I retrieved data on institutional ownership, managerial ownership, board composition and board independence from Execucomp, Thomson Reuters and BoardEx respectively. I use 10-K filings to create additional measure of financial constraint. To this effect, I parse ''Liquidity and Capital Resources'' subsection for all electronically filed 10-Ks for the period 1996 -- 2015. This approach will be discussed in detail later.
\paragraph{}
For the purposes of analyzing zero-leverage puzzle, difference between book and market leverage is not significant, since for unlevered firms numerator in both formulas is equal to zero. Thus further analysis uses only book leverage. Zero-leverage firm is defined as a firm, which has neither current nor long-term debt in the given year, while positive-leverage firms is defined as a firm, having non-zero current or long-term debt. Book leverage ratio of firm i in year t is defined as follows:
\begin{equation}
BL_{it} = (DLTT_{it}+DLC_{it})/(AT_{it}) ,
\end{equation}
where DLTT is the value of long-term debt and DLC is the value or current debt (debt with maturity below one year). This definition is compatible to recent papers (Strebulaev, 2012; Lemmon, Roberts and Zender, 2008; Graham and Leary, 2011). 
\paragraph{}
Appendix A provides an overview of all variables as well as detailed description of their calculation. The sample includes only US firms with positive total assets and excludes regulated firms (Standard Industry Classification codes 4900 -- 4999) as well as financial companies (SIC codes 6000 -- 6999). To make results robust to characteristics of small firms, sample includes only companies with total book value of assets more than \$10 million in 2015 dollars. To adjust for inflation all nominal variables, I use Consumer Price Index (CPI) data from Federal Reserve Bank of St. Louis. After excluding duplicated firm-year observations there are 104688 firm-year observations left. To mitigate effect of outliers, all variables are trimmed at 1\% and 99\% percentile levels. 
\paragraph{}
Table 1 presents dynamics of zero-leverage policy for the sample period. From 1996 to 2015 number of zero-leverage firms as a fraction of all public firms increased from 14.7\% to 22\%, showing recently increased significance of zero-leverage phenomenon. 2000 -- 2005 period accounts for the most rapid growth of zero-leverage firms percentage, when fraction of zero-leverage firms surged from 16\% to 23\%. Despite the fact that fraction of zero-leverage firms decreased by 2\% after the recent crisis, there is clear increasing long-term trend in zero-leverage policy. Zero-leverage companies accounts for 19.5\% of the companies in the whole sample. 
\paragraph{}
Tables 2, 3 and 4 reports distribution of zero-leverage firms among broad and narrow industries. Broad industry is defined based on 2-digit SIC code, while 3-digit SIC code is used to determine narrow industries.  There is some heterogeneity in occurrence of zero-leverage policy among broad industries. Construction as well as Transportation, Communication and Utilities have the lowest percentage of zero-leverage firms, which is 5.5\% and 7.4\% respectively. Services is the only broad industry, where fraction of zero-leverage firms is sufficiently higher than in the whole sample (28.4\% compared to 19.5\%). In spite of obvious differences of propensity to use zero-leverage policy among broad industries, unveiled distinctions are not large enough to explain all the variation in the sample, since fraction of zero-leverage companies is below 30\% in all broad industries.
\paragraph{}
Tables 3 and 4 describe heterogeneity in usage of zero-leverage policy among narrow industries. Educational services, Apparel and Accessory Stores, and Non-classifiable Establishments have the highest percentage of zero-leverage firms -- 32.6\%, 31.8\% and 31.6\% respectively. The list of 10 industries with the highest fraction of zero-leverage companies consists of different types of services as well as manufacturing industries. Interesting fact is that even in case of narrow industries there are no industries with very high proportion of zero-leverage companies, since the ‘most unlevered’ industry has only 32.6\% debt-free companies compared to 19.5\% in the whole sample. Local passenger transit is the only industry where all the firms have positive leverage. Other industries with the lowest percentage of zero-leverage firms include Automotive Dealers, Tobacco products, Petroleum and Coal, Auto Services, General Building Contractors, and Water Transportation. Fraction of debt-free firms is below 3\% in all aforementioned industries.
\paragraph{}
Uncovered differences between narrow industries can be explained using basic trade-off theory and cost of financial distress as the main distinctive feature of narrow industries. Services has the highest percentage of zero-leverage firms, since due to high relative importance of intangible assets, they lose relatively more value in case of financial distress compared to transportation companies, which are less prone to use zero-leverage policy.
\paragraph{}
Table 5 and Figure 1 provide insight into mean profitability of zero-leverage firms as well as positive-leverage firms. Within all the sample period, levered firms always were more profitable than zero-leverage firms. Starting from 1998 zero-leverage firms always on average had negative profitability, while  positive-leverage companies usually were profitable. Figure 1 illustrates that within 1996 -- 2015 profitabilities of zero-leverage and positive-leverage firms evolved in the similar way, but profitability of zero-leverage firms exhibites much higher volatility. Another interesting fact is that profitability of zero-leverage firms increased insignificantly during recession and then exhibited sudden decrease during recovery and expansion period. 
\paragraph{}
Table 6 presents summary statistic of zero-leverage firms and positive-leverage firms. To test statistical significance of differences in means of variables, Table 6 reports results of T-test as well. Zero-leverage firms on average are smaller, younger and less profitable (by means of profitability, returns on assets and returns on equity). These results are consistent with previous studies (Devos, 2008; Byoun, 2012; Strebulaev, 2013). Zero-leverage firms have less tangible assets and hold more cash than firms with positive leverage. These differences are extremely significant with absolute value of t-statistics being above 150. Obtained results are in accord with intuition, since tangible assets are the most convenient collateral for bank loans, while cash holdings are good source of liquidity which could be raised through borrowings otherwise. Zero-leverage firms have higher market valuation, captured by market-to-book ratio, price-to-sales ratio and Tobin`s Q. Interestingly enough, while having lower profitability, zero-leverage firms pay higher dividends than positive-leverage firms. One of possible explanation of this phenomenon is using dividends as substitution for debt payment in order to discipline managers and mitigate shareholder-management agency conflicts. Zero-leverage firms have higher R\&D expenses, but lower capital expenditures. On average, zero-leverage firms grow faster, have lower internal cash flow, issue more equity and have much higher volatility of profitability and cash flows. 
\paragraph{}
Many reported results are expected from extensive literature on leverage and its determinants (Lemmon, Roberts, Zender, 2008), but some evidences, such as lower profitability, cash flows and capital expenditures of zero-leverage firms are rather unexpected. As to profitability of zero-leverage firms compared to profitability of levered firms, the situation significantly changed over time. Although this timeframe is beyond sample, used in the current paper, zero-leverage firms had higher profitability compared to levered firms before 1980, while after 1980 positive-leverage companies have been more profitable than zero-levered firms with this gap constantly increasing. 
\paragraph{}
Table 7 reports results of propensity-score matching, where each zero-leverage firm was matched to three levered firms in the same 3-digit SIC industry, fiscal year and with total assets within 66\% - 150\% of its total assets. Summary statistics of zero-leverage firms and their matched levered peers differs significantly from previous results. Main reason for this difference is that size of the firm is extremely important variable, which is likely to influence all other variable. Therefore, when comparing zero-leverage firms to unmatched positive-leverage firms, the fact that zero-leverage firms are smaller than positive-leverage firms will influence the results, since in fact small firms are compared to large firms. Matching by size lets eliminate this channel and argue that the differences, which are left, are mostly attributable to leverage. 
\paragraph{}
After matching by fiscal year, narrow industry and size, zero-leverage firms are still younger, but difference becomes less significant. Probably it is caused by close correlation between size and age. The most interesting result is that after matching zero-leverage firms are more profitable (in terms of profitability, ROA and ROE) than their matched levered peers are. Zero-leverage firms have less tangible assets, hold more cash, pay higher dividends, grow faster and have higher valuations. While unlevered firms still have lower capital expenditures, difference in R\&D expenses becomes insignificant. Differences in mean standard deviations of profitability and cash flow also become less significant.
\paragraph{}
Although main objective of this section is to report basic evidence from the data, it is possible to draw some preliminary conclusions as to validity of the main explanations of the zero-leverage puzzle. Smaller size and age, lower tangibility and profitability (in case without matching) of zero-leverage firms are consistent with financial constraints explanation, although higher asset growth of zero-leverage firms is a minor contradiction to this theory. Financial flexibility explanation is not consistent with significantly higher R\&D expenses of zero-leverage firms. Summary statistics provides evidence for high market valuation explanation too, since zero-leverage firms have higher market valuation ratios. 
\paragraph{}
Table 8 reports evidences on transitivity of zero-leverage policy. While 19.5\% of all firms in the sample are zero-leverage firms, among entering firms 26.4\% are all-equity firms. This significant difference shows that fraction of zero-leverage firms is much higher for firms at the moment they enter Compustat. At each year, zero-leverage firm becomes levered with probability 19.9\%, while probability that levered firm becomes zero-leverage firm is only 3.4\%. These probabilities let us model system of firms, randomly moving between zero leverage and positive leverage, as discrete time Markov Chain. Using these transition probabilities, implied long-term distribution of firms must be 14.8\% and 85.2\%, which is significantly different from empirically observable one. Therefore, data shows that fraction of all-equity firms among entering firms is higher than the one among all firms in the sample, which in turn is higher than long-term fraction of unlevered firms, implied by transition probabilities. The most evident explanation of this phenomenon is that aging zero-leverage firms slowly become positive-leverage firms. Alternative explanation suggests that instead of transferring to positive-leverage firms, zero-leverage firms are more likely to delist. Since probability of delisting is 8.5\% for zero-leverage firm compared to 9.8\% for positive-leverage firm, results suggest that high fraction of zero-leverage firm in the data is driven by new firms, which are more likely to enter unlevered. 

\section{Empirical results} \label{sec:Model}

\subsection{5.1	Depletion of post-IPO cash reserves}
The first obvious test of post-IPO cash depletion explanation is to look whether post-IPO zero-leverage firms issued equity. Since this explanation suggests that young zero-leverage firms stay unlevered as long as they have enough cash reserves left after IPO, these firms should not issue equity as well as debt. However, in the sample of 4601 zero-leverage firm-year observations, 60.11\% of firm-year observation indicated positive equity issue. Therefore, this basic test implies that explanation of zero-leverage puzzle from the point of post-IPO cash depletion is not consistent with the data.
\paragraph{}
In order to test Hypothesis 1, I create variable After-IPO Cash Ratio, equal to ratio of current cash reserves to cash reserves immediately after IPO. Thus, the theory predicts that young zero-leverage firm gradually depletes cash holdings after IPO and After-IPO Cash Ratio will be gradually decreasing. Therefore, zero-leverage firms issue debt when After-IPO Cash Ratio is low, and we expect negative coefficient in regression of levering of zero-leverage firms on After-IPO Cash Ratio.
\paragraph{}
Table 9 reports results of regression of levering of zero-leverage firms on After-IPO Cash Ratio. Second and third model specifications include industry fixed effects at 2-digit and 3-digit SIC levels respectively. All three regression models estimate significantly positive coefficients on After-IPO Cash Ratio, implying then levering events of zero-leverage firms are associated with high current cash holdings comparing to immediate after-IPO cash reserves. These results falsify Hypothesis 1 and together with unexpectedly high equity issuance of young zero-leverage firms imply that we should reject explanation of zero-leverage puzzle from the point of view of depletion of post-IPO cash reserves.

\subsection{5.1	Financial constraints}
In order to test financial constraints theory, I use the following measures of financial constraints:
\begin{enumerate}
  \item Financial constraints from textual analysis of 10-K filings.
  \item SA-Index.
  \item Cash flow sensitivity of cash
\end{enumerate}
Appendix B describes in detail measuring financial constraints from 10-K annual filings. I use simplified approach, used by Hoberg and Maksimovic (2014) to perform textual analysis and compute financial constraints variable. 
\paragraph{}
Table 10 reports results of textual analysis of 59 958 10-K filings. Only 10-K, 10-KSB, 10-K405 and 10-KSB40 were analyzed for non-financial companies. For each fiscal year, fraction of firms, which appear to be financially constrained, is 7 -- 19\%. Appendix B provides some examples of 10-K filings, where financial constraints were identified. For the purposes of further analysis, I take log of variable “Constraints”.
\paragraph{}
SA-Index is another measure of financial constraints (Hadlock and Pierce, 2010). SA-Index is calculated as follows:
\begin{equation}
SA\ index = (-0.737*Size) + (0.043*Size^2) - (0.040*Age),
\end{equation}
where size was trimmed at \$5B and age was trimmed at 37 years. Higher values of SA-Index indicate higher level of financial constraint. The motivation behind functional form of SA-Index is that young and small firms are financially constrained more often, while dependence of constraint on size is non-linear.
\paragraph{}
Cash-flow sensitivity of cash is a measure of financial constraint, based on assumption that more financially constrained firms tend to save cash out of cash flows, since their cash flows are likely to be the only source of liquidity. Since firm can not be sure as to future cash flows, it saves cash from current cash flows to ensure against future cash flow disruptions. Almeida (2004) show that financially constrained firms usually save 5-6\% of cash flows in cash, while unconstrained firms do not save any statistically significant percentage of cash flows. To calculate this measure of financial constraints, I run the same regression as Almeida (2004):
\begin{equation}
\Delta {CashFlow}_{it} = alpha_{0} + alpha_{1}*CashFlow_{it} + alpha_{2}*MB_{it} + alpha_{3}*Size_{it}. 
\end{equation}
Firms for which coefficient on Cash flow was at least 0.05, were classified as financially constrained. This approach identifies 8352 firms out of 41223 as financially constrained. 
Table 11 presents results of test of Hypothesis 2A. When using measure of financial constraints, derived from 10-K filings, regression of zero-leverage status on logarithm of financial constraints estimates positive coefficient, implying that zero-leverage status is associated with higher values of financial constraints. After adding 2-digit and 3-digit industry fixed effects, coefficients decrease substantially, but remain statistically significant. Therefore, under measure of financial constraints from 10-Ks, Hypothesis 2A on zero-leverage firms being financially constrained is accepted.
\paragraph{}
Using SA-index as a measure of financial constraint gives similar results with even higher statistical significance. Since higher SA-index means higher level of financial constraint, positive and statistically significant coefficient in regression of zero-leverage status on SA-index suggests that zero-leverage status is associated with financial constraint. Adding traditional variables in leverage regressions as extra covariates and including industry fixed effects does not change results sufficiently. Coefficient on SA-index remains positive and very significant, implying that using SA-index as a measure of financial constraint provides strong evidence in favor of Hypothesis 2A.
\paragraph{}
Finally, cash flow sensitivity of cash as a measure of financial constraint returns similar results, as shown in the last three columns in Table 11. In regression of zero leverage status on financial constraint, identified from cash flow sensitivity of cash, coefficient on financial constraint dummy variable is positive and statistically significant. Controlling for traditional variables in leverage regressions and industry fixed effects decreases value of coefficient more than by half, but it remains significantly positive. Thus, all three measures of financial constraint provide strong evidence in favor of Hypothesis 2A, and thus financial constraint explanation of zero-leverage puzzle.

\subsection{Financial flexibility}
First obvious test of financial flexibility explanation of zero-leverage puzzle is to examine transitivity of zero-leverage status. Table 12 reports result of transitivity analysis. Since the sample includes 20 years of data, the largest possible number of consecutive years with zero leverage is 20 years. Financial flexibility theory predicts that for most firms zero leverage will be short-term phenomenon and unlevered firms will issue debt within couple of years after becoming all-equity firms. Table 12 reports that 50.7\% firms stay unlevered for 2 years or less and 63.5\% firms stay unlevered for less or equal to 3 years. These results show that behavior of half of firms is generally consistent with financial flexibility explanation, while the other half does not utilise zero-leverage policy as a short-term measure. So these results provide mixed evidence for financial flexibility theory.
\paragraph{}
Table 13 presents results of test of Hypothesis 3A. Hypothesis 3A entails that immediately after levering a firm increases investment or acquisition. Looking at R\&D expenses, no increase after levering takes place. Furthermore, there is statistically significant decrease in R\&D 1 year after levering. This is exacerbated by the fact that immediately after levering R\&D decreases too, though decrease is statistically insignificant. However, these two fact together imply that 1 and two years after levering R\&D expenses are significantly smaller than prior to levering. 
\paragraph{}
Dynamics of capital expenditures near levering looks completely different. It is difficult to answer a question whether capital expenditures or R\&D expenses are better variables to test financial flexibility theory, since in capital-intensive industries a role of capital expenditures is much higher than in less capital-intensive industries. One year before levering there is statistically significant increase in capital expenditures (from 4.4\% to 5.3\%), followed by another statistically significant increase immediately after levering (from 5.3\% to 6.2\%). During the next two years, capital expenditures gradually revert to its starting values. Significant increase in capital expenditures 1 year before levering may look suspicious, but financial flexibility theory entails that levering is planned decision of a firm, so firm may start to increase capex, already knowing that it will raise funds by issuing debt soon. So these results are consistent with predictions of financial flexibility theory.
\paragraph{}
Analysis of acquisition activities near levering provides strong evidence in favor of Hypothesis 3A. During all years before levering acquisition levels are small and never exceed 0.5\%. In the years of levering dramatic increase in acquisitions is registered, with acquisitions amounting to 6.3\% of total assets. For all subsequent years acquisition activities decrease substantially but still remain significantly higher than before levering. Therefore analysis of capital expenditures and acquisitions strongly accepts hypothesis 3A, while analysis of R\&D expenses does not provide evidence in favor of Hypothesis 3A.

\subsection{High market valuation}
Table 14 reports results of testing Hypothesis 4A regarding predictions of market valuation theory as to zero-leverage firms. Fisrt part of Hypothesis 4A entails that zero-leverage firms have decreasing market valuation prior to levering. In fact it is this decrease which makes them issue debt. Table 14 shows that neither market-to-book ration nor Tobin`s Q exhibit any significant decrease before levering: both variables are increasing, but this increase is not statistically significant. Price/sales ratio 2 years prior to levering is significantly higher than 3 year prior to levering, while 1 years before levering price/sales ratio is again higher comparing to previous year, the increase being even more statistically significant. Therefore, this test provides evidence against first part of Hypothesis 4A. 
\paragraph{}
Second part of Hypothesis 4A predicts that decreasing valuations of zero-leverage firms make them stop issuing equity and finally make them decide to lever. Table 14 reports that for all years before levering zero-leverage firms are issuing significantly more equity than levered firms. Thus results provide evidence against second part of Hypothesis 4A as well. Ultimately, tests of Hypothesis 4A give clear evidence against this hypothesis and against high market valuation explanation of zero-leverage puzzle in general. 

\subsection{Managerial entrenchment}
Hypothesis 5 states that firms with stronger corporate governance will be less likely to be unlevered. I use board independence and institutional ownership as proxies for quality of institutional governance. Furthermore, I use plausibly exogenous source of treatment: delisting shock. This variable calculates a change in percentage of firms, which were delisted due to mergers in the same 2-digit SIC industry in the previous year. The motivation behind this is a a notion that unusually high takeover activity in some industry should scare managers of all firms in the industry and make them worry as to possibility of being acquired. Poor management is known as one the most frequent reasons for takeover, so, to decrease probability of hostile takeover, managers will attempt to decrease suboptimal value-destroying activities. Under managerial entrenchment theory, zero-leverage policy is suboptimal, so it can cause hostile takeover.
\paragraph{}
Table 15 reports results of test for managerial entrenchment explanation of zero-leverage puzzle. Under hypothesis 5, higher board independence, institutional ownership and delisting shock should cause lower probability of zero-leverage policy, so coefficients on all variables should be negative. In regression of zero-leverage dummy variable on only these three covariates board independence have statistically significantly positive coefficient, while institutional ownership have negative statistically significant coefficient, as expected. For all possible model specifications, coefficient on delisting shock is not statistically significant. Second, third and fourth columns in Table 15 reports results of adding traditional capital structure variables, 2-digit and 3-digit fixed effects. All these regressions estimate significantly positive coefficients on board independence and institutional ownership. Therefore, results of the test of managerial entrenchment explanation clearly falsify Hypothesis 5 and suggest that managerial entrenchment theory is not valid explanation for zero-leverage puzzle.

\subsection{Multiple hypotheses}
Table 16 reports results of tests of first parts of Hypotheses 2A, 3A and 4A. In 2007 23\% unlevered firms issued debt, in 2008 only 20.9\% firms with zero leverage become levered. While this decrease from 2007 to 2008 is not statistically significant, dramatic decrease from 20.9\% in 2008 to 13.5\% in 2009 is obvious and statistically significant. Therefore, financial crisis clearly negatively affected probability of levering of zero-leverage firms. Thus we can see that this test falsifies Hypotheses 3A and 4A. Since high market valuation explanation (Hypothesis 4A) had already evidences against, at this point I dismiss this explanation of zero-leverage puzzle. Significant decrease in probability of levering for zero-leverage firms is evidence in favor of financial constraints explanation.
\paragraph{}
Second parts of Hypotheses 2A and 3A are not mutually exclusive and relate to changes of variables near transition from zero-leverage firms to positive-leverage firm. Table 17 presents evidence on Hypothesis 2A. Here I utilise difference-in-difference estimator with financial crisis as a source of exogenous shock. Thus effects of treatment and zero-leverage status are captured by coefficients on these variables, while effect of financial crisis on zero-leverage firms compared to its effect on positive-leverage firms is captured by Treatment*Zero-leverage coefficient. All indicators of performance (profitability, ROA, ROE, stock returns, market share) decreased for zero-leverage firms compared to positive-leverage firms after crisis, though this decrease was statistically significant only for ROA and returns. So this test provides mixed evidence in favor of financial constraints explanation.
\paragraph{}
Table 18 reports findings on Hypothesis 3A (financial flexibility). As expected, capital expenses, R\&D and acquisitions decreased after the crisis. While R\&D expenses of zero-leverage firms were not affected more or less than those of positive-leverage firms, capital expenditures and acquisitions for zero-levered firms increased dramatically compared to positive-leverage forms. These results provide evidence in favor of second part of Hypothesis 3A (financial flexibility).

\section{Conclusion}
This paper attempts to determine which theories of capital structure can explain zero-leverage behavior. Previous research offered numerous explanations of this phenomenon, including managerial entrenchment, financial constraints, financial flexibility and high equity market valuations. In order to find a theoretical explanation, most consistent with the data, I test predictions, generated by each of the aforementioned explanations. 
\paragraph{}
Results of the analysis clearly reject explanations of zero-leverage behavior from the point of view of weak corporate governance, since firms with higher board independence and institutional ownership are more likely to be unlevered. Tests falsify a hypothesis, resulting from an explanation of zero-leverage puzzle with high equity market valuation, since prior to levering, zero-leverage firms do not suffer from declining market valuations and continue to issue large amounts of equity. 
\paragraph{}
The paper finds strong evidence for both the financial constraints explanation and the financial flexibility explanation. Zero-leverage firms are more financially constrained than positive-leverage firms by three different measures of financial constraints and dramatically increase capital expenditures as well as acquisitions immediately after levering. To discriminate between the financial constraints explanation and financial flexibility theory, I use the recent financial crisis as an exogenous shock to the economic environment. Since a crisis is unlikely to make previously financially constrained zero-leverage firms less financially constrained, the percentage of zero-leverage firms, deciding to issue debt, should decrease. The financial flexibility explanation predicts that zero-leverage firms will use their stockpiled debt capacity to issue debt in order to avoid cutting investments or acquiring distressed competitors. A dramatic decrease in the proportion of levering zero-leverage firms in 2009 provides clear and statistically significant evidence in favor of the financial constraints explanation and against the financial flexibility explanation.
\paragraph{}
 This paper shows that borrowing constraints explain zero-leverage behavior better than more complicated explanations, such as weak corporate governance or high equity valuations. My results suggest that financial constraints are the key driving force behind zero-leverage policy and contribute to the research of the zero-leverage and low-leverage puzzle. Furthermore, I unveil the role of financial constraints in zero-leverage behavior and calls upon theoretical studies of capital structure, accounting for borrowing constraints. On the one hand, financial constraints are important determinants of zero-leverage policy but are unlikely to be a key factor in the more general theory of capital structure. On the other hand, traditional capital structure theories, such as trade-off theory, explain capital structure decisions relatively well except in cases of zero and ultralow leverage.  Therefore theoretical research, incorporating borrowing constraints as an important friction into the dynamic trade-off model look specifically promising.

\section{References:}
\begin{enumerate}
\item Almeida, H., Campello, M., Weisbach, M.S., 2004. The cash flow sensitivity of cash. J. Finance 59, 1777–1804.
\item	Berger, P.G., Ofek, E., Yermack, D.L., 1997. Managerial entrenchment and capital structure decisions. J. Finance 52, 1411–1438.
\item	Bessler, W., Drobetz, W., Haller, R., and Meier, I. (2013). The international zero-leverage phenomenon. Journal of CorporateFinance, 23 (12), 196-221.
\item	Boroshko T., Kokoreva M., 2014. The phenomenon of debt absence in the capital structure of companies in the emerging capital markets. Working paper.
\item	Byoun, S. and Xu, Z. (2013) Why do some firms go debt free?, Asia-Pacific Journal of Financial Studies, 42, 1–38. 
\item	Dang, V., 2011. An empirical analysis of zero-leverage firms: evidence from the UK. Unpublished working paper. University of Manchester, Manchester Business School, Manchester, England.
\item	DeAngelo, H., DeAngelo, L., 2007. Capital structure, payout policy, and financial flexibility. University of Southern California, Working Paper.
\item	Devos Erik, Upinder Dhillon, Murali Jagannathan, Srinivasan Krishnamurthy, “Why Are Firms Unlevered?,” Journal of Corporate Finance 18, no. 3 (June 2012): 664–682
\item	Faulkender, Michael, and Mitchell A. Petersen, 2006, Does the source of capital affect capital structure?, Review of Financial Studies 19, 45-79.
\item	Fischer, E., Heinkel, R., Zechner, J., 1989. Optimal dynamic capital structure choice: theory and tests. Journal of Finance 44, 19–40.
\item	Graham, J., 2000. How big are the tax benefits of debt?, Journal of Finance 55, 1901-1942.
\item	Graham, J.G., Harvey, C.R., 2001. The theory and practice of corporate finance: evidence from the field. Journal of Financial Economics 60, 187–243.
\item	Graham, J., Leary, M., 2011. A review of empirical capital structure research and directions for the future. Annual Review of Financial Economics 3, 309–345.
\item   Hoberg, Gerard and Vojislav Maksimovic, 2015, “Redefining Financial Constraints: A Text-Based Analysis,” Review of Financial Studies, 28(5), 1312-1352.
\item	Jensen, M.C., 1986. Agency costs of free cash flow, corporate finance, and takeovers. Am. Econ. Rev. 76, 323–329.
\item	Hadlock, C.J., Pierce, J.R., 2010. New evidence on measuring financial constraints: moving beyond the KZ index. Rev. Financ. Stud. 23, 1909–1940.
\item	Lee, H., Moon, G., 2011. The long-run equity performance of zero-leverage firms. Managerial Finance 37, 872–889.
\item	Lee, H., Moon, G., Waggle D., 2014. The effect of debt capacity on the long-term stock returns of debt-free firms, Applied Economics, 47:4, 333-345
\item	Lemmon, M., Roberts, M., Zender, J., 2008. Back to the beginning: persistence and the cross-section of corporate capital structure. Journal of Finance 63, 1575–1608.
\item	Minton, B., Wruck, K., 2001. Financial conservatism: evidence on capital structure from low leverage firms. Unpublished working paper. Ohio State University, Columbus, OH.
\item	Strebulaev, I., 2007. Do tests of capital structure theory mean what they say? Journal of Finance 62, 2633–2671.
\item	Strebulaev Ilya A. and Baozhong Yang, “The Mystery of Zero-Leverage Firms,” Journal of Financial Economics 109, no. 1 (July 2013): 1–23.

\end{enumerate}

\section*{Appendix A}
\subsection*{Definition of variables}
\begin{table}[H]
\begin{tabular}{ l l}
Size & log(Total Assets) \\
Age	& number of years the firm is in Compustat \\ [0.5ex]
Profitability &  $\frac{EBIT}{Total Assets}$  \\ [0.5ex]
Tangibility	 &  $\frac{Property,\ Plant\ and Equipment}{Total\ Assets}$ \\ [0.5ex]
Cash & $\frac{Cash\ and\ Short-Term\ Investments}{Total\ Assets}$ \\ [0.5ex]
R\&D & $\frac{Research\ and\ Development\ Expense}{Total\ Assets}$ \\ [0.5ex]
MB (Market-to-Book) & $\frac{Total\ Assets - Total\ Common\ Equity + Market\ Value}{Total\ Assets}$ \\ [0.5ex]
P/S (Price/Sales) & $\frac{(Common\ Shares\ Outstanding)(Price\ Close)}{Sales}$ \\ [0.5ex]
ROA & $\frac{Net\ Income}{Total\ Assets}$ \\ [0.5ex]
ROE & $\frac{Net\ Income}{Total\ Common\ Equity}$ \\ [0.5ex]
Asset Growth & $\frac{Total\ Assets_{t}}{Total\ Assets_{t-1}}$ - 1 \\ [0.5ex]
Dividends & $\frac{Dividends\ Common}{Total\ Assets}$ \\ [0.5ex]
Capex & $\frac{Capital\ Expenditures}{Assets\ Total}$ \\ [0.5ex]
Cash Flow & $\frac{Income\ Before\ Extraordinary\ Items + Depreciation\ and\ Amortization}{Total\ Assets}$ \\ [0.5ex]
Net Debt Issuance & $\frac{Tota\l Debt_{t}}{Total\ Debt_{t-1}}$ - 1\\ [0.5ex]
Equity Issuance & $\frac{Sale\ of\ Common\ and\ Preferred\ Stock - Purchase\ of\ Common\ and\ Preferred\ Stock}{Total\ Assets}$ \\ [0.5ex]
Tobin`s Q & $\frac{Total\ Assets + Market\ Value + Common\ Equity + Deferred\ Taxes}{Total\ Assets}$ \\ [0.5ex]
Acquisitions & $\frac{Acquisitions}{Total\ Assets}$ \\ [0.5ex]
Book Leverage & $\frac{LongTerm Debt + ShortTerm Debt}{Total\ Assets}$ \\ [0.5ex]
After-IPO Cash Ratio & $\frac{Cash_{it}}{Cash_{i0}}$ \\ [0.5ex]
SA-index & $ (0.737*Size) + (0.043*Size^2) - (0.040*Age)$ \\ [0.5ex]
Board\ Independence & $\frac{Number of external directors}{Board Size}$ \\ [0.5ex]
Delisting\ Shock & $\max{(\frac{Fraction\ of\ delistings\ due\ to\ merger\ in\ 2-digit\ SIC\ industry_t}{Fraction\ of\ delistings\ due\ to\ merger\ in\ 2-digit\ SIC\ industry_{t-1}} - 1, 0)}$ \\ [0.5ex]

\end{tabular}
\end{table}

\newpage
\section*{Appendix B}
\subsection*{Textual analysis of 10-K filings for financial constraints}
I parse “Liquidity and Capital Resourses” subsection from 10-K, 10-K405, 10-KSB and 10-KSB40 filings for the period 1996 – 2015. This subsection contains remarks regarding financial condition and liquidity sources as well as intentions with regard to future capital market interactions. \\
To create measure of financial constraints I use approach similar to one, used by Hoberg and Maksimovic (2014). This approach assumes that the firms, which discuss in “Liquidity and Capital Resources” delaying investments are financially constrained. Since firms discuss liquidity issues in this subsection, it follows logically that firms decrease investments due to liquidity concerns. \\
In order to identify delaying investments I use composition of words from 2 lists. To find firms which directly report potential delay in investments, I search “Liquidity and Capital Resources” subsection for occurrences of words from list 1, identify their locations and then search for occurrences of words from list 2 in close neighbourhood to locations of words from list 1. \\
List 1 consists of verbs, synonymous to “delay”, while list 2 contains nouns and phrases, standing for processes, suffering from delays due to liquidity issues. Note that “*” denotes wildcard. \\ \\
\textbf{List 1: delay*, abandon*, eliminate*, curtail*, postpone*, curb*, suspend*, defer*, discontinue*, cut* back, scale* down} \\ \\
\textbf{List 2: construction*,  expansion*, acquisition*, restructuring, project*, research, development, exploration, product*, expenditure*, manufactur*, entry*, renovat*, growth, activities, activity, capital improvement*, capital spend*, capital proj*, commercial release*, business plan, transmitter deployment, opening restaurants.} \\ \\
After word from list 1 is identified, I search for words from list 2 within 10 words after found word from list 1.

\subsection*{Examples of 10-Ks with identified financial constraints}
\subsubsection*{10-K for 2014 fiscal year of Paratek Pharmaceuticals:}
“… Prior to the Merger and recapitalization in October 2014 we were subject to significant liquidity constraints. During 2014 and 2013 we significantly \textbf{curtailed} our \textbf{research} and \textbf{development} and other operating activities as we worked within financial constraints…”
\subsubsection*{10-K for 2014 fiscal year of Nanoantibiotics:}
“… However, there can be no assurance that we will be able to obtain required funding.  If we are unsuccessful in securing funding from any of these sources, we will \textbf{defer}, reduce or \textbf{eliminate} certain planned \textbf{expenditures} in our research protocols.  If we do not have sufficient funds to continue operations, we could be required to seek bankruptcy protection or other alternatives that could result in our stockholders losing some or all of their investment in us…”
\subsubsection*{10-K for 2013 fiscal year by HKN Inc.:}
“…There can be no assurances that, if needed, we will be able to raise adequate funds from these or other sources or enter into licensing, partnering or other arrangements to advance our business goals. Our inability to raise such funds or enter into such arrangements, if needed, could have a material adverse effect on our ability to develop our products. Also, we have the ability to \textbf{curtail} discretionary \textbf{spending}, including some \textbf{research} and \textbf{development activities}, if required to conserve cash. Because of our long-term capital requirements, we may seek to access the public equity market whenever conditions are favorable, even if we do not have an immediate need for additional capital at that time…”
\subsubsection*{10-K for 2013 fiscal year by Nanoantibiotics:}
“…We continue to focus on improving our operating efficiency to increase operating cash flows, since we are unlikely to have access to public markets. Our actions have included implementing operational expense reduction initiatives, \textbf{delaying} or \textbf{eliminating} certain capital \textbf{spending} and \textbf{research} and \textbf{development projects}, focusing on timely customer collections and re-negotiating longer payment terms with our vendors. We hope that we will be able to satisfy our cash requirements for at least the next twelve months with the liquidity provided by our existing cash, cash equivalents and marketable securities…”
\subsubsection*{10-K for 2013 fiscal year by Coronado Biosciences:}
“…Adequate additional funding, particularly subsequent to the negative results from our TRUST-I clinical trial, may not be available to us on acceptable terms or at all. If adequate funds are not available to us when needed, we may be required to \textbf{delay, curtail} or \textbf{eliminate} one or more of our \textbf{research} and \textbf{development} programs and, potentially, \textbf{delay} our \textbf{growth} strategy…”

\pagebreak

\begin{table}[!tbp]
\caption*{Table 1: Fraction of zero-leverage firms over time}
\begin{center}
\begin{tabular}{rllr}
\hline\hline
\multicolumn{1}{c}{Year}&\multicolumn{1}{c}{Zero leverage}&\multicolumn{1}{c}{Normal leverage}&\multicolumn{1}{c}{Obs}\tabularnewline
\hline
$1996$&14.67\%&85.33\%&$5795$\tabularnewline
$1997$&14.89\%&85.11\%&$5701$\tabularnewline
$1998$&14.97\%&85.03\%&$5830$\tabularnewline
$1999$&15.28\%&84.72\%&$5721$\tabularnewline
$2000$&15.99\%&84.01\%&$5360$\tabularnewline
$2001$&16.57\%&83.43\%&$4907$\tabularnewline
$2002$&17.72\%&82.28\%&$4656$\tabularnewline
$2003$&19.73\%&80.27\%&$4531$\tabularnewline
$2004$&21.67\%&78.33\%&$4411$\tabularnewline
$2005$&22.92\%&77.08\%&$4384$\tabularnewline
$2006$&22.94\%&77.06\%&$4264$\tabularnewline
$2007$&22.92\%&77.08\%&$4131$\tabularnewline
$2008$&21.28\%&78.72\%&$4050$\tabularnewline
$2009$&22.14\%&77.86\%&$3849$\tabularnewline
$2010$&23.55\%&76.45\%&$3702$\tabularnewline
$2011$&23.45\%&76.55\%&$3731$\tabularnewline
$2012$&22.74\%&77.26\%&$3822$\tabularnewline
$2013$&22.84\%&77.16\%&$3910$\tabularnewline
$2014$&22.31\%&77.69\%&$3801$\tabularnewline
$2015$&21.92\%&78.08\%&$3112$\tabularnewline
\hline
\end{tabular}\end{center}
\end{table}

\begin{table}[!tbp]
\caption*{Table 2: Fraction of zero-leverage firms in broad industries}
\begin{center}
\begin{tabular}{lllr}
\hline\hline
\multicolumn{1}{l}{2-digit SIC Industry}&\multicolumn{1}{c}{Zero}&\multicolumn{1}{c}{Normal}&\multicolumn{1}{c}{Observations}\tabularnewline
\hline
Agriculture, Forestry and Fishing&9.61\%&90.39\%&$   461$\tabularnewline
Mining&14.59\%&85.41\%&$  6762$\tabularnewline
Construction&5.48\%&94.52\%&$  1419$\tabularnewline
Manufacturing&20.29\%&79.71\%&$ 50722$\tabularnewline
Transportation, Communication and Utilities&7.42\%&92.58\%&$  9038$\tabularnewline
Wholesale Trade&10.2\%&89.8\%&$  4155$\tabularnewline
Retail Trade&14.37\%&85.63\%&$  7571$\tabularnewline
Services&28.37\%&71.63\%&$ 23160$\tabularnewline
All industries&19.49\%&80.51\%&$104688$\tabularnewline
\hline
\end{tabular}\end{center}
\end{table}

\begin{table}[!tbp]
\caption*{Table 3: Industries with the highest fraction of zero-leverage firms}
\begin{center}
\begin{tabular}{llr}
\hline\hline
\multicolumn{1}{l}{}&\multicolumn{1}{c}{3-digit SIC Industry}&\multicolumn{1}{c}{Fraction}\tabularnewline
\hline
1&Educational Services&$32.6\%$\tabularnewline
2&Apparel \& Accessory Stores&$31.8\%$\tabularnewline
3&Non-Classifiable Establishments&$31.6\%$\tabularnewline
4&Business Services&$27.7\%$\tabularnewline
5&Instruments \& Related Products&$24.9\%$\tabularnewline
6&Leather&$23.9\%$\tabularnewline
7&Pipelines, Except Natural Gas&$23.7\%$\tabularnewline
8&Transportation Services&$23.1\%$\tabularnewline
9&Chemical \& Allied products&$22.5\%$\tabularnewline
10&Engineering \& Management Services&$22.5\%$\tabularnewline
\hline
\end{tabular}\end{center}
\end{table}

\begin{table}[!tbp]
\caption* {Table 4: Industries with the lowest fracion of zero-leverage firms}
\begin{center}
\begin{tabular}{llr}
\hline\hline
\multicolumn{1}{l}{}&\multicolumn{1}{c}{3-digit SIC Industry}&\multicolumn{1}{c}{Fraction}\tabularnewline
\hline
1&Local Passenger Transit&$0.0\%$\tabularnewline
2&Automative Dealers \& Service Stations&$1.8\%$\tabularnewline
3&Tobacco Products&$1.8\%$\tabularnewline
4&Petroleum \& Coal&$2.0\%$\tabularnewline
5&Auto Services&$2.2\%$\tabularnewline
6&General Building Contractors&$2.5\%$\tabularnewline
7&Water Transportation&$2.7\%$\tabularnewline
8&Paper \& Allied Products&$2.9\%$\tabularnewline
9&Transportation by Air&$3.8\%$\tabularnewline
10&Social Services&$3.9\%$\tabularnewline
\hline
\end{tabular}\end{center}
\end{table}

\begin{table}[!tbp]
\caption*{Table 5: Mean profitability of different leverage groups over time}
\begin{center}
\begin{tabular}{rrr}
\hline\hline
\multicolumn{1}{c}{Years}&\multicolumn{1}{c}{Zero}&\multicolumn{1}{c}{Normal}\tabularnewline
\hline
$1996$&$ 0.9\%$&$ 3.1\%$\tabularnewline
$1997$&$ 0.5\%$&$ 2.0\%$\tabularnewline
$1998$&$-5.3\%$&$-1.5\%$\tabularnewline
$1999$&$-6.1\%$&$-1.1\%$\tabularnewline
$2000$&$-10.8\%$&$-1.1\%$\tabularnewline
$2001$&$-14.8\%$&$-2.6\%$\tabularnewline
$2002$&$-9.1\%$&$ 0.0\%$\tabularnewline
$2003$&$-4.7\%$&$ 1.6\%$\tabularnewline
$2004$&$-3.3\%$&$ 3.4\%$\tabularnewline
$2005$&$-3.7\%$&$ 2.4\%$\tabularnewline
$2006$&$-4.4\%$&$ 2.3\%$\tabularnewline
$2007$&$-4.2\%$&$ 1.8\%$\tabularnewline
$2008$&$-5.1\%$&$ 0.7\%$\tabularnewline
$2009$&$-2.7\%$&$ 1.3\%$\tabularnewline
$2010$&$-1.3\%$&$ 4.1\%$\tabularnewline
$2011$&$-3.6\%$&$ 3.3\%$\tabularnewline
$2012$&$-6.7\%$&$ 1.5\%$\tabularnewline
$2013$&$-8.4\%$&$ 0.9\%$\tabularnewline
$2014$&$-11.0\%$&$ 0.7\%$\tabularnewline
$2015$&$-17.5\%$&$-1.1\%$\tabularnewline
\hline
\end{tabular}\end{center}
\end{table}

\begin{landscape}\begin{table}[!tbp]
\caption*{Table 6: Summary statistics}
\begin{center}
\begin{tabular}{lrrrrrr}
\hline\hline
\multicolumn{1}{l}{Variable}&\multicolumn{2}{c}{Zero-leverage}&\multicolumn{2}{c}{Positive leverage}&\multicolumn{2}{c}{T-test}\tabularnewline
\hline\hline
\multicolumn{1}{l}{}&\multicolumn{1}{c}{Mean}&\multicolumn{1}{c}{Median}&\multicolumn{1}{c}{Mean}&\multicolumn{1}{c}{Median}&\multicolumn{1}{c}{Diff in means}&\multicolumn{1}{c}{T-stat}\tabularnewline
\hline
Size&$ 4.594$&$4.589$&$ 6.072$&$ 6.147$&$-1.478$&$-120.190$\tabularnewline
Age&$11.562$&$9.000$&$16.118$&$11.000$&$-4.557$&$ -45.524$\tabularnewline
Profitability&$-0.013$&$0.023$&$ 0.049$&$ 0.062$&$-0.062$&$ -46.259$\tabularnewline
Tangibility&$ 0.095$&$0.067$&$ 0.260$&$ 0.216$&$-0.165$&$-161.476$\tabularnewline
Cash&$ 0.421$&$0.395$&$ 0.087$&$ 0.051$&$ 0.334$&$ 168.966$\tabularnewline
R\&D&$ 0.138$&$0.110$&$ 0.059$&$ 0.029$&$ 0.079$&$  69.297$\tabularnewline
MB&$ 2.103$&$1.675$&$ 1.401$&$ 1.241$&$ 0.701$&$  58.718$\tabularnewline
P/S&$ 3.500$&$2.630$&$ 0.978$&$ 0.835$&$ 2.522$&$  98.716$\tabularnewline
ROA&$-0.029$&$0.022$&$ 0.003$&$ 0.022$&$-0.032$&$ -24.032$\tabularnewline
ROE&$-0.032$&$0.032$&$ 0.022$&$ 0.072$&$-0.054$&$ -27.356$\tabularnewline
Asset growth&$ 0.069$&$0.040$&$ 0.058$&$ 0.026$&$ 0.012$&$   6.208$\tabularnewline
Dividends&$ 0.009$&$0.000$&$ 0.008$&$ 0.000$&$ 0.001$&$   5.165$\tabularnewline
Capex&$ 0.028$&$0.021$&$ 0.045$&$ 0.035$&$-0.017$&$ -68.447$\tabularnewline
Cash Flow&$-0.011$&$0.055$&$ 0.043$&$ 0.068$&$-0.054$&$ -31.865$\tabularnewline
Net debt issuance&$-0.004$&$0.000$&$ 0.026$&$ 0.005$&$-0.030$&$ -72.829$\tabularnewline
Equity issuance&$ 0.042$&$0.003$&$ 0.004$&$ 0.000$&$ 0.037$&$  37.451$\tabularnewline
Profitability SD&$ 0.145$&$0.118$&$ 0.074$&$ 0.054$&$ 0.071$&$  73.737$\tabularnewline
CF SD&$ 0.160$&$0.125$&$ 0.086$&$ 0.060$&$ 0.074$&$  59.824$\tabularnewline
Tobin`s Q&$ 3.595$&$3.174$&$ 2.132$&$ 2.012$&$ 1.463$&$ 119.016$\tabularnewline
\hline
\end{tabular}\end{center}
\end{table}\end{landscape}

\begin{landscape}\begin{table}[!tbp]
\caption*{Table 7: Summary statistics with matching}
\begin{center}
\begin{tabular}{lrrrrrr}
\hline\hline
\multicolumn{1}{l}{Variable}&\multicolumn{2}{c}{Zero-leverage}&\multicolumn{2}{c}{Positive leverage}&\multicolumn{2}{c}{T-test}\tabularnewline
\hline\hline
\multicolumn{1}{l}{}&\multicolumn{1}{c}{Mean}&\multicolumn{1}{c}{Median}&\multicolumn{1}{c}{Mean}&\multicolumn{1}{c}{Median}&\multicolumn{1}{c}{Diff(means)}&\multicolumn{1}{c}{T-stat}\tabularnewline
\hline
Age&$ 9.953$&$7.000$&$10.225$&$ 7.000$&$-0.272$&$ -2.886$\tabularnewline
Profitability&$-0.020$&$0.018$&$-0.052$&$ 0.003$&$ 0.032$&$ 19.330$\tabularnewline
Tangibility&$ 0.132$&$0.073$&$ 0.208$&$ 0.136$&$-0.077$&$-49.699$\tabularnewline
Cash&$ 0.456$&$0.429$&$ 0.222$&$ 0.129$&$ 0.234$&$ 95.843$\tabularnewline
R\&D&$ 0.146$&$0.108$&$ 0.150$&$ 0.090$&$-0.003$&$ -2.061$\tabularnewline
MB&$ 1.884$&$1.596$&$ 1.537$&$ 1.343$&$ 0.347$&$ 31.120$\tabularnewline
P/S&$ 8.794$&$2.914$&$ 2.340$&$ 1.201$&$ 6.454$&$ 33.944$\tabularnewline
ROA&$-0.037$&$0.018$&$-0.110$&$-0.034$&$ 0.073$&$ 41.754$\tabularnewline
ROE&$-0.042$&$0.028$&$-0.209$&$-0.025$&$ 0.167$&$ 53.657$\tabularnewline
Asset growth&$ 0.169$&$0.046$&$ 0.135$&$ 0.025$&$ 0.034$&$  7.424$\tabularnewline
Dividends&$ 0.006$&$0.000$&$ 0.001$&$ 0.000$&$ 0.004$&$ 33.097$\tabularnewline
Capex&$ 0.040$&$0.025$&$ 0.050$&$ 0.030$&$-0.009$&$-21.699$\tabularnewline
Cash Flow&$ 0.001$&$0.052$&$-0.050$&$ 0.020$&$ 0.051$&$ 29.920$\tabularnewline
Net debt issuance&$-0.009$&$0.000$&$ 0.040$&$ 0.011$&$-0.048$&$-67.593$\tabularnewline
Equity issuance&$ 0.109$&$0.004$&$ 0.079$&$ 0.003$&$ 0.030$&$ 12.683$\tabularnewline
Profitability SD&$ 0.159$&$0.122$&$ 0.152$&$ 0.110$&$ 0.007$&$  4.829$\tabularnewline
CF SD&$ 0.159$&$0.122$&$ 0.161$&$ 0.125$&$-0.003$&$ -1.906$\tabularnewline
Tobin`s Q&$ 3.366$&$3.091$&$ 2.289$&$ 2.176$&$ 1.078$&$ 91.623$\tabularnewline
\hline
\end{tabular}\end{center}
\end{table}\end{landscape}

\begin{table}[!tbp]
\caption*{Table 8: Transitivity of zero-leverage policy}
\begin{center}
\begin{tabular}{lll}
\hline\hline
\multicolumn{1}{l}{}&\multicolumn{1}{c}{Zero Leverage}&\multicolumn{1}{c}{Positive leverage}\tabularnewline
\hline
Fraction of firms in data&19.5\%&80.5\%\tabularnewline
Fraction of entering firms&26.4\%&73.6\%\tabularnewline
Probability of changing leverage group&19.9\%&3.46\%\tabularnewline
Implied long-term fraction of firms&14.8\%&85.2\%\tabularnewline
Probability of delisting&8.5\%&9.8\%\tabularnewline
\hline
\end{tabular}\end{center}
\end{table}

\clearpage

\begin{table}[!htbp] \centering  
  \label{} 
\caption*{Table 9: Post-IPO cash depletion}
\begin{tabular}{@{\extracolsep{5pt}}lccc} 
\\[-1.8ex]\hline 
\hline \\[-1.8ex] 
 & \multicolumn{3}{c}{\textit{Dependent variable:}} \\ 
\cline{2-4} 
\\[-1.8ex] & \multicolumn{3}{c}{Levering dummy variable} \\ 
\\[-1.8ex] & (1) & (2) & (3)\\ 
\hline \\[-1.8ex] 
 After-IPO Cash Ratio & 0.0004$^{**}$ & 0.0003$^{**}$ & 0.0003$^{**}$ \\ 
  & (0.0002) & (0.0002) & (0.0002) \\ 
  & & & \\ 
 Constant & $-$1.654$^{***}$ & $-$2.486$^{**}$ & $-$2.486$^{**}$ \\ 
  & (0.040) & (1.041) & (1.041) \\ 
  & & & \\ 
\hline \\[-1.8ex] 
Industry Fixed Effects & No & 2-digit & 3-digit \\
Observations & 4,601 & 4,601 & 4,601 \\ 
Log Likelihood & $-$2,028.230 & $-$1,977.937 & $-$1,942.693 \\ 
Akaike Inf. Crit. & 4,060.459 & 4,059.874 & 4,149.387 \\ 
\hline 
\hline \\[-1.8ex] 
\textit{Note:}  & \multicolumn{3}{r}{$^{*}$p$<$0.1; $^{**}$p$<$0.05; $^{***}$p$<$0.01} \\ 
\end{tabular} 
\end{table} 

\begin{table}[!tbp]
\caption*{Table 10: Financial Constraints from 10-K filings}
\begin{center}
\begin{tabular}{lrr}
\hline\hline
\multicolumn{1}{l}{Constraints}&\multicolumn{1}{c}{Number of filings}&\multicolumn{1}{c}{Fraction, \%}\tabularnewline
\hline
0&$51535$&$85.952$\tabularnewline
1&$ 3725$&$ 6.213$\tabularnewline
2&$ 2074$&$ 3.459$\tabularnewline
3&$  873$&$ 1.456$\tabularnewline
4&$  466$&$ 0.777$\tabularnewline
5&$  306$&$ 0.510$\tabularnewline
6&$  231$&$ 0.385$\tabularnewline
7&$  148$&$ 0.247$\tabularnewline
8&$  119$&$ 0.198$\tabularnewline
9&$  103$&$ 0.172$\tabularnewline
10&$   78$&$ 0.130$\tabularnewline
11&$   48$&$ 0.080$\tabularnewline
12&$   54$&$ 0.090$\tabularnewline
13&$   36$&$ 0.060$\tabularnewline
14&$   24$&$ 0.040$\tabularnewline
15&$   28$&$ 0.047$\tabularnewline
16&$   20$&$ 0.033$\tabularnewline
17&$   19$&$ 0.032$\tabularnewline
18&$   19$&$ 0.032$\tabularnewline
19&$    8$&$ 0.013$\tabularnewline
20&$   12$&$ 0.020$\tabularnewline
21&$   10$&$ 0.017$\tabularnewline
22&$    3$&$ 0.005$\tabularnewline
23&$    4$&$ 0.007$\tabularnewline
24&$    4$&$ 0.007$\tabularnewline
25&$    2$&$ 0.003$\tabularnewline
26&$    1$&$ 0.002$\tabularnewline
27&$    2$&$ 0.003$\tabularnewline
29&$    1$&$ 0.002$\tabularnewline
32&$    1$&$ 0.002$\tabularnewline
34&$    2$&$ 0.003$\tabularnewline
35&$    1$&$ 0.002$\tabularnewline
47&$    1$&$ 0.002$\tabularnewline
\hline
Total&$ 59958 $&$ 100.000$\tabularnewline
\hline
\end{tabular}\end{center}
\end{table}

\newgeometry{left=2cm, right=1cm, top=1cm, bottom=1.5cm} 
\begin{landscape}
\begin{table}[!htbp] \centering 
\footnotesize
\caption*{Table 11: Financial Constraints}
\begin{tabular}{@{\extracolsep{5pt}}lccccccccc} 
\\[-1.8ex]\hline 
\hline \\[-1.8ex] 
 & \multicolumn{9}{c}{\textit{Dependent variable:}} \\ 
\cline{2-10} 
\\[-1.8ex] & \multicolumn{9}{c}{Zero leverage dummy variable} \\ 
\\[-1.8ex] & (1) & (2) & (3) & (4) & (5) & (6) & (7) & (8) & (9)\\ 
\hline \\[-1.8ex] 
 log (10-K\ Constraints) & 0.244$^{***}$ & 0.124$^{***}$ & 0.051$^{**}$ &  &  &  &  &  &  \\ 
  & (0.022) & (0.022) & (0.023) &  &  &  &  &  &  \\ 
  & & & & & & & & & \\ 
 SA\ -\ Index &  &  &  & 0.539$^{***}$ & 0.378$^{***}$ & 0.381$^{***}$ &  &  &  \\ 
  &  &  &  & (0.016) & (0.028) & (0.029) &  &  &  \\ 
  & & & & & & & & & \\ 
 Profitability &  &  &  &  & 1.079$^{***}$ & 1.090$^{***}$ &  & 0.946$^{***}$ & 1.322$^{***}$ \\ 
  &  &  &  &  & (0.077) & (0.079) &  & (0.164) & (0.171) \\ 
  & & & & & & & & & \\ 
 Tangibility &  &  &  &  & $-$1.888$^{***}$ & $-$1.822$^{***}$ &  & $-$3.738$^{***}$ & $-$3.636$^{***}$ \\ 
  &  &  &  &  & (0.132) & (0.141) &  & (0.131) & (0.142) \\ 
  & & & & & & & & & \\ 
 MB &  &  &  &  & $-$0.002 & $-$0.003 &  & 0.179$^{***}$ & 0.146$^{***}$ \\ 
  &  &  &  &  & (0.009) & (0.009) &  & (0.012) & (0.012) \\ 
  & & & & & & & & & \\ 
 Cash &  &  &  &  & 4.690$^{***}$ & 4.608$^{***}$ &  &  &  \\ 
  &  &  &  &  & (0.086) & (0.090) &  &  &  \\ 
  & & & & & & & & & \\ 
 Dividends &  &  &  &  & 5.072$^{***}$ & 5.096$^{***}$ &  &  &  \\ 
  &  &  &  &  & (0.463) & (0.484) &  &  &  \\ 
  & & & & & & & & & \\ 
 CF\ SD &  &  &  &  & $-$0.281$^{***}$ & $-$0.321$^{***}$ &  &  &  \\ 
  &  &  &  &  & (0.091) & (0.092) &  &  &  \\ 
  & & & & & & & & & \\ 
 Cash\ Sensitivity\ Constraints &  &  &  &  &  &  & 0.972$^{***}$ & 0.421$^{***}$ & 0.384$^{***}$ \\ 
  &  &  &  &  &  &  & (0.030) & (0.037) & (0.037) \\ 
  & & & & & & & & & \\ 
 Age &  &  &  &  &  &  &  & 0.003$^{**}$ & 0.006$^{***}$ \\ 
  &  &  &  &  &  &  &  & (0.002) & (0.002) \\ 
  & & & & & & & & & \\ 
 Size &  &  &  &  &  &  &  & $-$0.277$^{***}$ & $-$0.286$^{***}$ \\ 
  &  &  &  &  &  &  &  & (0.011) & (0.011) \\ 
  & & & & & & & & & \\ 
 Constant & $-$1.584$^{***}$ & $-$2.402$^{***}$ & $-$2.395$^{***}$ & 0.140$^{***}$ & $-$1.597$^{***}$ & $-$1.581$^{***}$ & $-$1.873$^{***}$ & 0.052 & 0.060 \\ 
  & (0.012) & (0.290) & (0.290) & (0.048) & (0.477) & (0.476) & (0.017) & (0.395) & (0.395) \\ 
  & & & & & & & & & \\ 
\hline \\[-1.8ex] 
Industry Fixed Effects & No & 2-digit & 3-digit & No & 2-digit & 3-digit & No & 2-digit & 3-digit \\
Observations & 59,958 & 59,958 & 59,958 & 57,292 & 35,211 & 35,211 & 38,826 & 31,190 & 31,190 \\ 
Log Likelihood & $-$27,878.210 & $-$26,034.700 & $-$24,795.690 & $-$26,688.520 & $-$13,103.910 & $-$12,682.080 & $-$16,927.690 & $-$12,084.430 & $-$11,545.720 \\ 
Akaike Inf. Crit. & 55,760.420 & 52,201.400 & 50,083.390 & 53,381.030 & 26,347.810 & 25,860.160 & 33,859.380 & 24,310.870 & 23,589.440 \\ 
\hline 
\hline \\[-1.8ex] 
\textit{Note:}  & \multicolumn{9}{r}{$^{*}$p$<$0.1; $^{**}$p$<$0.05; $^{***}$p$<$0.01} \\ 
\end{tabular} 
\end{table}
\end{landscape}
\restoregeometry

\begin{table}[!tbp]
\normalsize
\caption*{Table 12: Transitivity of zero-leverage policy}
\begin{center}
\begin{tabular}{lrr}
\hline\hline
\multicolumn{1}{l}{Consecutive years}&\multicolumn{1}{c}{Number of firms}&\multicolumn{1}{c}{Fraction, \%}\tabularnewline
\hline
1&$6061$&$31.231$\tabularnewline
2&$3799$&$19.575$\tabularnewline
3&$2513$&$12.949$\tabularnewline
4&$1764$&$ 9.090$\tabularnewline
5&$1273$&$ 6.559$\tabularnewline
6&$ 949$&$ 4.890$\tabularnewline
7&$ 727$&$ 3.746$\tabularnewline
8&$ 528$&$ 2.721$\tabularnewline
9&$ 403$&$ 2.077$\tabularnewline
10&$ 313$&$ 1.613$\tabularnewline
11&$ 256$&$ 1.319$\tabularnewline
12&$ 189$&$ 0.974$\tabularnewline
13&$ 151$&$ 0.778$\tabularnewline
14&$ 121$&$ 0.623$\tabularnewline
15&$  96$&$ 0.495$\tabularnewline
16&$  72$&$ 0.371$\tabularnewline
17&$  61$&$ 0.314$\tabularnewline
18&$  55$&$ 0.281$\tabularnewline
19&$  51$&$ 0.260$\tabularnewline
20&$  41$&$ 0.209$\tabularnewline
\hline
\end{tabular}\end{center}
\end{table}

\begin{table}[!tbp]
\caption*{Table 13. Financial Flexibility Hypothesis: \\ Mean variables around transition from zero leverage to positive leverage}
\begin{center}
\begin{tabular}{lrrrrrr}
\hline\hline
\multicolumn{1}{l}{}&\multicolumn{1}{c}{-3 years}&\multicolumn{1}{c}{-2 years}&\multicolumn{1}{c}{-1 year}&\multicolumn{1}{c}{0 year}&\multicolumn{1}{c}{+1 year}&\multicolumn{1}{c}{+2 years}\tabularnewline
\hline
R\&D &$ 0.120$&$ 0.121$&$ 0.127$&$ 0.123$&$ 0.101$&$ 0.092$\tabularnewline
R\&D t-stat&$ $&$0.15$&$0.96$&$ -0.66$&$-3.35$&$-1.364$\tabularnewline
Capex &$ 0.045$&$ 0.044$&$ 0.053$&$ 0.062$&$ 0.051$&$ 0.042$\tabularnewline
Capex t-stat&$ $&$-0.29$&$ 3.39$&$ 2.93$&$ -3.57$&$ -1.62$\tabularnewline
Acquisitions &$ 0.004$&$ 0.004$&$ 0.005$&$ 0.063$&$ 0.015$&$ 0.018$\tabularnewline
Acquisitions t-stat&$ $&$ 0.44$&$ 1.21$&$14.88$&$ -11.65$&$ 0.98$\tabularnewline
\hline
\end{tabular}\end{center}
\end{table}

\begin{table}[!tbp]
\caption*{Table 14. High Market Valuation Hypothesis:\\
Mean Variables around Transition Event\footnotemark\\}
\begin{center}
\begin{tabular}{lrrrrrr}
\hline\hline
\multicolumn{1}{l}{}&\multicolumn{1}{c}{-3 years}&\multicolumn{1}{c}{-2 years}&\multicolumn{1}{c}{-1 year}&\multicolumn{1}{c}{0 year}&\multicolumn{1}{c}{+1 year}&\multicolumn{1}{c}{+2 years}\tabularnewline
\hline
MB levering&$ 1.875$&$ 1.895$&$ 1.979$&$  1.624$&$ 1.619$&$ 1.479$\tabularnewline
MB t-stat&$ $&$ 0.25$&$ 1.16$&$ -5.934$&$-0.11$&$-3.06$\tabularnewline
P/S levering&$ 3.335$&$ 3.958$&$ 5.284$&$  2.918$&$ 2.418$&$ 2.040$\tabularnewline
P/S t-stat&$ $&$2.98$&$4.70$&$ -9.33$&$-4.13$&$ 3.54$\tabularnewline
Tobin`s Q levering&$ 3.298$&$ 3.352$&$ 3.419$&$  2.623$&$ 2.575$&$ 2.441$\tabularnewline
Tobin`s Q t-stat&$ $&$0.63$&$0.86$&$ -12.53$&$-0.92$&$-2.57$\tabularnewline
Equity iss. levering&$ 0.118$&$ 0.144$&$ 0.152$&$  0.072$&$ 0.080$&$ 0.052$\tabularnewline
Equity iss. unlevering&$ 0.061$&$ 0.062$&$ 0.107$&$  0.215$&$ 0.055$&$ 0.035$\tabularnewline
Equity iss. t-stat&$ 4.895$&$ 7.127$&$ 3.622$&$-10.942$&$ 2.982$&$ 3.636$\tabularnewline
Debt iss. levering&$-0.006$&$-0.003$&$ 0.000$&$  0.215$&$ 0.026$&$ 0.018$\tabularnewline
Debt iss. unlevering&$-0.013$&$-0.013$&$-0.026$&$ -0.158$&$ 0.000$&$ 0.013$\tabularnewline
Debt iss. t-stat&$ 1.460$&$ 2.217$&$ 6.264$&$ 42.549$&$ 7.050$&$ 1.044$\tabularnewline
\hline
\end{tabular}\end{center}
\end{table}
\footnotetext{`levering` means transition of the firms from zero-leverage firm to positive-leverage firm.\\ `unlevering' stands for transition from positive-leverage firm to zero-leverage firm}

\clearpage

\begin{table}[!htbp] \centering 
  \caption*{Table 15: Managerial Entrenchment} 
  \label{} 
\begin{tabular}{@{\extracolsep{5pt}}lcccc} 
\\[-1.8ex]\hline 
\hline \\[-1.8ex] 
 & \multicolumn{4}{c}{\textit{Dependent variable:}} \\ 
\cline{2-5} 
\\[-1.8ex] & \multicolumn{4}{c}{Zero-leverage dummy variable} \\ 
\\[-1.8ex] & (1) & (2) & (3) & (4)\\ 
\hline \\[-1.8ex] 
Board\ Independence & 0.803$^{***}$ & 0.297$^{**}$ & 0.423$^{***}$ & 0.468$^{***}$ \\ 
  & (0.104) & (0.147) & (0.154) & (0.160) \\ 
  & & & & \\ 
 Institutional\ Ownership & $-$0.799$^{***}$ & 1.231$^{***}$ & 1.058$^{***}$ & 1.035$^{***}$ \\ 
  & (0.050) & (0.086) & (0.091) & (0.094) \\ 
  & & & & \\ 
 Delisting\ Shock & 0.267 & $-$0.042 & $-$0.053 & $-$0.008 \\ 
  & (0.396) & (0.539) & (0.648) & (0.661) \\ 
  & & & & \\ 
 Age &  & $-$0.004$^{**}$ & $-$0.001 & $-$0.002 \\ 
  &  & (0.002) & (0.002) & (0.002) \\ 
  & & & & \\ 
 Size &  & $-$0.571$^{***}$ & $-$0.552$^{***}$ & $-$0.551$^{***}$ \\ 
  &  & (0.017) & (0.018) & (0.019) \\ 
  & & & & \\ 
 Profitability &  & 2.311$^{***}$ & 1.660$^{***}$ & 1.699$^{***}$ \\ 
  &  & (0.111) & (0.115) & (0.117) \\ 
  & & & & \\ 
 Tangibility &  & $-$0.932$^{***}$ & $-$1.128$^{***}$ & $-$1.110$^{***}$ \\ 
  &  & (0.119) & (0.166) & (0.184) \\ 
  & & & & \\ 
 MB &  & 0.002 & 0.022 & 0.027$^{*}$ \\ 
  &  & (0.013) & (0.014) & (0.014) \\ 
  & & & & \\ 
 Cash &  & 4.176$^{***}$ & 4.638$^{***}$ & 4.455$^{***}$ \\ 
  &  & (0.101) & (0.113) & (0.119) \\ 
  & & & & \\ 
 Dividends &  & 4.539$^{***}$ & 4.879$^{***}$ & 4.436$^{***}$ \\ 
  &  & (0.494) & (0.520) & (0.532) \\ 
  & & & & \\ 
 CF\_SD &  & $-$0.268$^{**}$ & $-$0.228$^{*}$ & $-$0.307$^{**}$ \\ 
  &  & (0.119) & (0.120) & (0.121) \\ 
  & & & & \\ 
 Constant & $-$1.511$^{***}$ & 0.119 & $-$1.267 & $-$1.250 \\ 
  & (0.073) & (0.131) & (1.030) & (1.031) \\ 
  & & & & \\ 
\hline \\[-1.8ex] 
Industry Fixed Effects & No & No & 2-digit SIC & 3-difit SIC \\
Observations & 31,054 & 26,366 & 26,366 & 26,366 \\ 
Log Likelihood & $-$15,453.070 & $-$9,620.751 & $-$9,016.584 & $-$8,596.348 \\ 
Akaike Inf. Crit. & 30,914.140 & 19,265.500 & 18,177.170 & 17,670.700 \\ 
\hline 
\hline \\[-1.8ex] 
\textit{Note:}  & \multicolumn{4}{r}{$^{*}$p$<$0.1; $^{**}$p$<$0.05; $^{***}$p$<$0.01} \\ 
\end{tabular} 
\end{table}

\begin{table}[!tbp]
\caption*{Table 16: Probability of zero-leevrage firm becoming positive-leverage firm}
\begin{center}
\begin{tabular}{lrrrrr}
\hline\hline
\multicolumn{1}{c}{Year}&\multicolumn{1}{p{3cm}}{Number of zero-levrage firms}&\multicolumn{1}{p{3.5cm}}{Number of levering zero-leverage firms}&\multicolumn{1}{p{5.25cm}}{Fraction of zero-leevrage firms, which become unlevered, \%}&\multicolumn{1}{p{2cm}}{X2-stat}\tabularnewline
\hline
$1996$&$699$&$182$&$26.0$&$$\tabularnewline
$1997$&$777$&$212$&$27.3$&$ 0.232$\tabularnewline
$1998$&$754$&$210$&$27.9$&$ 0.036$\tabularnewline
$1999$&$750$&$210$&$28.0$&$ 0.000$\tabularnewline
$2000$&$804$&$194$&$24.1$&$ 2.824$\tabularnewline
$2001$&$783$&$167$&$21.3$&$ 1.615$\tabularnewline
$2002$&$781$&$145$&$18.6$&$ 1.699$\tabularnewline
$2003$&$842$&$139$&$16.5$&$ 1.050$\tabularnewline
$2004$&$879$&$144$&$16.4$&$ 0.000$\tabularnewline
$2005$&$927$&$161$&$17.4$&$ 0.246$\tabularnewline
$2006$&$899$&$180$&$20.0$&$ 1.946$\tabularnewline
$2007$&$867$&$199$&$23.0$&$ 2.078$\tabularnewline
$2008$&$799$&$167$&$20.9$&$ 0.905$\tabularnewline
$2009$&$801$&$108$&$13.5$&$14.947$\tabularnewline
$2010$&$806$&$129$&$16.0$&$ 1.836$\tabularnewline
$2011$&$786$&$140$&$17.8$&$ 0.801$\tabularnewline
$2012$&$763$&$155$&$20.3$&$ 1.415$\tabularnewline
$2013$&$796$&$140$&$17.6$&$ 1.714$\tabularnewline
$2014$&$776$&$178$&$22.9$&$ 6.643$\tabularnewline
$2015$&$638$&$140$&$21.9$&$ 0.146$\tabularnewline
\hline
\end{tabular}\end{center}
\end{table}

\newgeometry{left=3cm, right=1cm, top=1cm, bottom=1.5cm} 
\begin{landscape}
\begin{table}[!htbp] \centering
\small
  \caption*{Table 17: Financial Crisis and Financial Constraints Hypothesis} 
  \label{} 
\begin{tabular}{@{\extracolsep{5pt}}lccccc} 
\\[-1.8ex]\hline 
\hline \\[-1.8ex] 
 & \multicolumn{5}{c}{\textit{Dependent variable:}} \\ 
\cline{2-6} 
\\[-1.8ex] & Profitability & ROA & ROE & Returns & Marketshare \\ 
\\[-1.8ex] & (1) & (2) & (3) & (4) & (5)\\ 
\hline \\[-1.8ex] 
 Treatment & $-$0.006 & 0.020$^{***}$ & 0.100$^{***}$ & 1.116$^{***}$ & 0.010$^{**}$ \\ 
  & (0.004) & (0.005) & (0.020) & (0.017) & (0.004) \\ 
  & & & & & \\ 
 Zero-leverage & $-$0.044$^{***}$ & $-$0.007 & 0.006$$ & 0.144$^{***}$ & $-$0.079$^{***}$ \\ 
  & (0.007) & (0.008) & (0.012) & (0.027) & (0.007) \\ 
  & & & & & \\ 
 Treatment*Zero-leverage & $-$0.004 & $-$0.007 & $-$0.107$^{**}$ & $-$0.154$^{***}$ & $-$0.007 \\ 
  & (0.010) & (0.012) & (0.046) & (0.039) & (0.010) \\ 
  & & & & & \\ 
 Constant & 0.036$^{***}$ & $-$0.044$^{***}$ & $-$0.124$^{***}$ & $-$0.803$^{***}$ & 0.094$^{***}$ \\ 
  & (0.003) & (0.004) & (0.014) & (0.012) & (0.003) \\ 
  & & & & & \\ 
\hline \\[-1.8ex] 
Observations & 7,415 & 7,415 & 7,415 & 6,659 & 7,461 \\ 
R$^{2}$ & 0.009 & 0.003 & 0.003 & 0.425 & 0.038 \\ 
Adjusted R$^{2}$ & 0.009 & 0.002 & 0.003 & 0.425 & 0.037 \\ 
Residual Std. Error & 0.171 (df = 7411) & 0.204 (df = 7411) & 0.780 (df = 7411) & 0.633 (df = 6655) & 0.167 (df = 7457) \\ 
F Statistic & 22.192$^{***}$ (df = 3; 7411) & 6.378$^{***}$ (df = 3; 7411) & 8.210$^{***}$ (df = 3; 7411) & 1,638.844$^{***}$ (df = 3; 6655) & 97.511$^{***}$ (df = 3; 7457) \\ 
\hline 
\hline \\[-1.8ex] 
\textit{Note:}  & \multicolumn{5}{r}{$^{*}$p$<$0.1; $^{**}$p$<$0.05; $^{***}$p$<$0.01} \\ 
\end{tabular} 
\end{table}
\end{landscape}

\newgeometry{left=1cm, right=1cm, top=2.5cm, bottom=3cm}
\begin{table}[!htbp] \centering 
  \caption*{Table 18: Financial Crisis and Financial Flexibility} 
  \label{} 
\begin{tabular}{@{\extracolsep{5pt}}lccc} 
\\[-1.8ex]\hline 
\hline \\[-1.8ex] 
 & \multicolumn{3}{c}{\textit{Dependent variable:}} \\ 
\cline{2-4} 
\\[-1.8ex] & Capex & R\&D & Acquisitions \\ 
\\[-1.8ex] & (1) & (2) & (3)\\ 
\hline \\[-1.8ex] 
 Treatment & $-$0.023$^{***}$ & $-$0.006 & $-$0.018$^{***}$ \\ 
  & (0.001) & (0.004) & (0.001) \\ 
  & & & \\ 
 Zero-leverage & $-$0.025$^{***}$ & 0.075$^{***}$ & $-$0.015$^{***}$ \\ 
  & (0.002) & (0.006) & (0.002) \\ 
  & & & \\ 
 Treatment*Zero-leverage & 0.010$^{***}$ & $-$0.010 & 0.020$^{***}$ \\ 
  & (0.003) & (0.008) & (0.003) \\ 
  & & & \\ 
 Constant & 0.067$^{***}$ & 0.063$^{***}$ & 0.026$^{***}$ \\ 
  & (0.001) & (0.003) & (0.001) \\ 
  & & & \\ 
\hline \\[-1.8ex] 
Observations & 7,417 & 4,523 & 7,125 \\ 
R$^{2}$ & 0.051 & 0.061 & 0.033 \\ 
Adjusted R$^{2}$ & 0.051 & 0.060 & 0.032 \\ 
Residual Std. Error & 0.058 (df = 7413) & 0.119 (df = 4519) & 0.045 (df = 7121) \\ 
F Statistic & 133.454$^{***}$ (df = 3; 7413) & 97.594$^{***}$ (df = 3; 4519) & 79.964$^{***}$ (df = 3; 7121) \\ 
\hline 
\hline \\[-1.8ex] 
\textit{Note:}  & \multicolumn{3}{r}{$^{*}$p$<$0.1; $^{**}$p$<$0.05; $^{***}$p$<$0.01} \\ 
\end{tabular} 
\end{table}

\end{document}